# Fundamental evaluation of the pressure gradient for lubrication flows in varying channels:

*Application to inertial Newtonian flow in a linear channel.*

**Panagiotis Sialmas, Kostas D. Housiadas**

*Department of Mathematics, University of the Aegean, Karlovassi, Samos, 83200, Greece*

**Abstract**

We derive general analytical expressions for the pressure gradient of a viscoelastic Oldroyd-B fluid in a symmetric channel with slowly varying geometry. Using classic lubrication theory and assuming isothermal, incompressible, and steady inertial flow, we present four fundamental methods for deriving these expressions: the first is based on the momentum balance, the second on the total force balance in the channel, the third on a higher moment of the momentum balance combined with the continuity equation, and the fourth on the mechanical energy of the flow. Additionally, we formally prove that the first and second methods yield identical expressions and highlight the significance of the exact analytical solution for the Oldroyd-B model along the channel walls. For a Newtonian fluid at the creeping flow limit the first and second methods give the well-known result that the pressure gradient along the channel equals the shear stress at the walls. The differences between the methods and the outcomes are also presented and discussed. Additionally, we derive a new set of lubrication equations based on the streamfunction, transformed coordinates that map the variable channel wall shape onto fixed ones, and new components of the viscoelastic extra-stress tensor. This formulation facilitates a much easier implementation of the continuity equation, the total mass balance, and the total force balance in the system, ensuring that non-physical or spurious solutions do not arise. Finally, we present in detail a limiting nonlinear case by considering the Newtonian inertial flow in a linearly varying channel. Through the implementation of our formulation, we provide strong evidence for the theoretical and numerical equivalence of the general expressions for the pressure gradient and the average pressure drop in the channel.

# 1. Introduction

The lubrication approximation—also known as the *slender channel approximation* or *thin gap approximation*—is an asymptotic method used to simplify the governing equations for a wide range of phenomena involving transport of mass, energy, and/or momentum. The first to conduct a lubrication analysis was Reynolds in 1886 [1], who studied a thin film of liquid separating a pair of solid objects, typically in relative motion. In his original work, Reynolds considered a simple Newtonian fluid and ignored fluid inertia, thus deriving a simple linear set of equations which was solved analytically to yield the solution for the velocity and pressure field.

The theoretical developments of Reynolds [1] can also be applied to a particularly interesting class of internal flows—pressure-driven flows of complex fluids, such as generalized Newtonian and viscoelastic fluids, in narrow and long tubes (planar channels or axisymmetric circular pipes), which are bounded by fixed walls. These pressure-driven flows are widely encountered in industrial processes, such as extrusion [2,3], and in applications such as microfluidic extensional rheometers [4] and devices for subcutaneous drug administration [5,6], among many others. In particular, contraction/expansion flows have attracted significant attention [7-9], as they are often used to determine important rheological properties of viscoelastic fluids—such as extensional viscosity—by relating the average pressure drop along the tube to the flow rate [10-17].

Much later than Reynolds' work, Williams [18,19] presented an analysis for deriving the lubrication equations for the laminar flow of a Newtonian fluid in 2D planar channels and 3D axisymmetric circular pipes, including fluid inertia. Moreover, Williams introduced mapped coordinates and the streamfunction to convert the continuity equation and two components of the momentum balance into a single nonlinear partial differential equation for the streamfunction. Williams observed that the final equation for the streamfunction was identical to Prandtl's boundary layer equation, but with different accompanying boundary conditions. He also focused on the existence of self-similar analytical solutions and found that similarity solutions exist only for linearly converging/diverging 2D planar channels and for circular pipes whose walls vary slowly and exponentially. Later, Bottler [20] numerically solved Williams' lubrication equation using the finite difference method to find non-similar solutions. Van Dyke [21] offered a detailed discussion and presented systematic approaches for deriving approximate (mainly analytical) solutions to the Navier-Stokes equations at the creeping flow limit, as well as for solving the lubrication equations with the inertia contribution, as the latter was included in the mathematical model by Williams [18,19].

The seminal work of Williams [18,19], followed by many scientists and researchers, focused not only on attempts to find similarity solutions to the lubrication equations but also on developing methods and techniques to find more general solutions [22-24]. It is worth mentioning that the similarity solutions for flows in tubes (planar channels or cylindrical pipes) satisfy only boundary conditions along the wall(s) of the channel (or boundary conditions at one of the walls and symmetry conditions in the midplane), or along the wall of the pipe, while boundary conditions at the inlet or outlet of the tube cannot be imposed or satisfied.

For internal confined flows in tubes, the essence of the lubrication approximation lies in the fact that when the maximum distance between the channel walls or the maximum radius of a circular pipe is substantially smaller than its length, the flow reaches a fully developed state along most of the tube. In contrast, flow rearrangements near the entrance and exit regions—resulting from specific flow conditions at the inlet and outlet cross-sections—can be safely ignored.

This fundamental idea—that in slender geometries a fully developed flow neglects entry and exit effects entirely—applies to steady, Newtonian, and laminar flows and leads to one of the most famous results in fluid mechanics: the Hagen-Poiseuille law, which relates the average pressure drop in a tube, $\Delta\Pi^*$, to the volumetric flow rate, $Q^*$; throughout the paper a star superscript denotes dimensional quantity. Specifically, for plane Poiseuille flow between two infinitely long parallel plates at a distance $h^*$, the relation is $\Delta\Pi^* = 12 Q^* \eta_s^* / h^{*3}$, while for Poiseuille flow in a long cylindrical pipe of constant cross-section with radius $h^*$, it is $\Delta\Pi^* = 8 Q^* \eta_s^* \ell^* / (\pi h^{*4})$ where $\eta_s^*$ is the constant viscosity of the fluid, and $\ell^*$ is the length of the pipe. It is now well accepted that the Hagen-Poiseuille law does not hold (i) near the tube entrance, (ii) near the tube exit, (iii) when the tube length is comparable to the wall spacing in the planar case or to the radius of the cross-section in the case of a circular pipe, and (iv) when the Reynolds number exceeds a critical threshold, leading to flow instabilities and eventually turbulence, which significantly increases the pressure drop required to maintain the same flow rate.

Precisely under those circumstances, the lubrication approximation breaks down. In other words, when entrance and exit effects influence the flow over the entire length of the tube, when the tube is short, or when the dimensionless numbers in the lubrication equations (such as the Reynolds number) exceed a critical threshold, the lubrication approximation is no longer adequate for providing reliable or accurate results. Worth noting here is that the entrance effects can be minimized when the inlet conditions for the field variables are consistent with similarity solution(s) of the lubrication equations.

Despite its limitations, the lubrication approximation has been extensively used over the years for modeling slow flows in confined and narrow geometries, thin fluid films [25-28], free surface flows [29], the motion of particles near surfaces [30,31], and flows in microchannels with known geometry [32-35], among many others. Nowadays, this method is considered a classic, serving as an invaluable and simple mathematical tool for modeling fluid flows.

It is also interesting that, for a long time after the work of Williams [18,19], the evaluation of the pressure gradient at each cross-section of the tube and the average pressure drop required to maintain a constant flow rate were completely overlooked. Sisavath *et al.* [36] recognized the importance of these quantities for engineering applications and brought this issue to attention. Since then, most researchers using the lubrication approximation have focused on the evaluation of the pressure gradient and average pressure drop.

The theoretical development of the governing equations at the classic lubrication limit, including the effect of viscoelasticity in Poiseuille (pressure-driven) or Couette (flows driven by the relative motion of the walls) flows, was much more challenging than the corresponding equations for a Newtonian viscous fluid with or without inertia. For confined flows in non-uniform tubes with solid walls, this analysis was first performed by Tichy in 1996 [37], using the Upper Convected Maxwell (UCM) model. Tichy's work has been followed by many others [38-43]. Additionally, theoretical investigations of the effect of viscoelasticity, augmented by numerical simulations, on contracting planar channels and circular pipes has also been carried out [44-52]. We emphasize, however, that the final lubrication equations for the constitutive model remain non-linear and are very similar to the original equations. Even at the creeping flow limit, obtaining numerical or approximate analytical solutions, or similarity solutions, remains a very difficult task. So far, similarity solutions have been derived by Sialmas & Housiadas [53] only for the steady, incompressible, and creeping flow of an Oldroyd-B fluid in hyperbolic 2D planar geometry or in hyperbolic 3D axisymmetric geometry. Like the similarity solutions previously derived for Newtonian fluids, these similarity solutions satisfy all the boundary conditions for the velocity components, as well as the exact analytical solution for the polymer extra-stress components along the wall(s) of the tube and on the midplane or the axis of symmetry of the tube. However, no inlet or outlet conditions are or can be satisfied, in full accordance with the fundamental idea of the lubrication approximation for slender tubes, under which the entry/exit effects are neglected.

Finally, we mention that the lubrication equations for an Oldroyd-B fluid contain two additional dimensionless parameters: the polymer viscosity ratio (which ranges from zero to one) and the Deborah number, which is the ratio of the polymer's relaxation time to the characteristic residence time of the fluid in the tube; see Section 2 below for their precise definitions. The magnitude of the polymer viscosity ratio does not significantly affect the main features of the lubrication equations or the results (at least qualitatively). However, the magnitude of the Deborah number requires special attention because a high Deborah number may induce significant changes to the flow due to the inlet or outlet conditions for the polymer extra-stresses (note that this does not apply to the similarity solution(s)). In this case, the inlet/outlet effects become significant, which contrasts with the main feature of the lubrication approximation, where inlet/outlet effects are neglected. This implies that when the Deborah number is sufficiently large, the lubrication approximation is no longer valid.

The main purpose of this work is to provide general theoretical expressions for the pressure gradient and the average pressure drop required to drive the flow with a constant volumetric flow rate by relaxing the creeping flow assumption as inertia can be significant in many circumstances [54]. These expressions are derived at the classic lubrication limit, based on first principles, using an integrated form of the momentum balance, the total force balance, a higher-order moment of the momentum balance combined with the continuity equation, and the total mechanical energy of the flow. Each method is presented separately, and the differences between the theoretical outcomes are highlighted and discussed. Additionally, a new set of lubrication equations is developed to ensure the exact satisfaction of the continuity equation, the total mass balance, and the total force balance by consistently incorporating the exact solution of the Oldroyd-B model along the channel walls.

We emphasize that we do not address issues regarding the inlet/outlet conditions for the polymer extra-stress tensor; some relevant information, however, can be found in [46,50-51]. The upper limit of the Deborah number that can be used under the lubrication approximation, as well as the existence of solutions of the Oldroyd-B model, are open issues that require additional consideration, new results, comparisons between different numerical and asymptotic approaches, and comparison with experimental data.

Finally, to illustrate the theoretical and numerical equivalence of the general expressions for the pressure gradient at any cross-section of the channel, along with the formulation in terms of mapped coordinates and the streamfunction, we present a limiting nonlinear case. Specifically, we consider the Newtonian inertial flow in a linearly varying channel and demonstrate that when the flow field is resolved with high accuracy, all the derived expressions for the pressure gradient yield the same results.

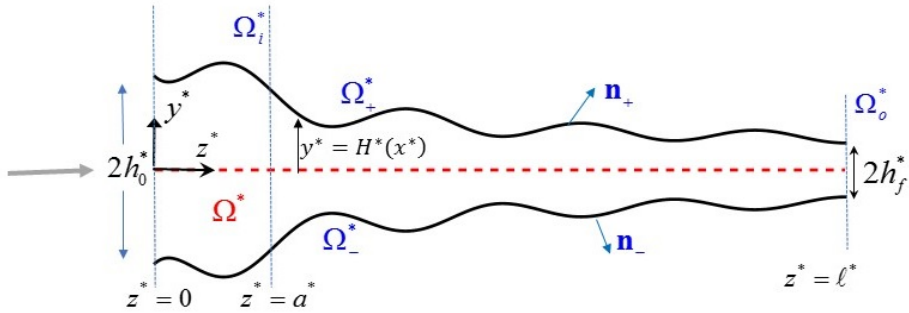

**Figure 1:** Flow geometry and coordinate system ($y^*, z^*$) for a symmetric channel. The flow domain is denoted by $\Omega^*$ and its boundary by $\partial\Omega^* = \Omega_i^* \cup \Omega_+^* \cup \Omega_-^* \cup \Omega_o^*$ where $\Omega_i^*$ is the cross-section at $z^* = a^*$ with unit normal vector $-\mathbf{e}_z$, $\Omega_o^*$ is the outlet cross-section at $z^* = \ell^*$ with unit normal vector $\mathbf{e}_z$, $\Omega_+^*$ is the upper wall surface with unit normal vector $\mathbf{n}_+$, and $\Omega_-^*$ is the lower wall surface with unit normal vector $\mathbf{n}_-$. The flow is from left to right, and the red dashed line denotes the midplane. A superscript start denotes a dimensional quantity.

## 2. Problem formulation

In Figure 1, a symmetric channel with respect to the midplane is depicted. The distance of the two walls at the inlet is $2h_0^*$ and at the outlet is $2h_f^*$, while its length is $\ell^*$. A constant volumetric flow rate, $Q^*$, and an incompressible Oldroyd-B fluid with constant mass density $\rho^*$ are considered. The fluid consists of a Newtonian solvent with constant shear viscosity $\eta_s^*$, and a polymeric material with constant shear viscosity $\eta_p^*$ and longest relaxation time $\lambda^*$. We use a Cartesian coordinate system ($z^*, y^*, x^*$) to describe the flow field, where $z^*$ is the main flow direction, $y^*$ is the vertical direction, and $x^*$ is the neutral direction; $\mathbf{e}_x, \mathbf{e}_z$ and $\mathbf{e}_y$ are the unit vectors in the $x^*$, $z^*$, and $y^*$ directions, respectively. Due to the symmetry with respect to the midplane, the walls of the channel are described by the shape function $H^* = H^*(z^*) > 0$ for $0 \leq z^* \leq \ell^*$, i.e. $y^* = H^*(z^*)$ for the upper wall and $y^* = -H^*(z^*)$ for the lower one. The velocity vector in the flow domain is denoted by $\mathbf{u}^* = V^*(y^*,z^*)\mathbf{e}_y + U^*(y^*,z^*)\mathbf{e}_z$ and the total pressure by $P^* = P^*(y^*,z^*)$.

Using the unit tensor $\mathbf{I}$, the rate of deformation tensor $\dot{\boldsymbol{\gamma}}^* = \nabla^*\mathbf{u}^* + (\nabla^*\mathbf{u}^*)^T$ where $\nabla^* \equiv \mathbf{e}_y(\partial/\partial y^*) + \mathbf{e}_z(\partial/\partial z^*)$ is the gradient operator, and the viscoelastic extra-stress tensor $\boldsymbol{\tau}^*$ we define the total momentum tensor per fluid volume:

$$\mathbf{T}^* := -\rho^* \mathbf{u}^* \mathbf{u}^* - P^*\mathbf{I} + \eta_s^* \dot{\boldsymbol{\gamma}}^* + \boldsymbol{\tau}^* \qquad (1)$$

Ignoring any external forces and torques, the equations that govern the flow in the channel are the mass and momentum balances, respectively:

$$\nabla^* \cdot \mathbf{u}^* = 0, \quad \nabla^* \cdot \mathbf{T}^* = \mathbf{0} \qquad (2a,b)$$

In order to determine $\boldsymbol{\tau}^*$ we use the Oldroyd-B model:

$$\boldsymbol{\tau}^* + \lambda^* \left( \mathbf{u}^* \cdot \nabla^* \boldsymbol{\tau}^* - \boldsymbol{\tau}^* \cdot \nabla^* \mathbf{v}^* - (\nabla^* \mathbf{v}^*)^T \cdot \boldsymbol{\tau}^* \right) = \eta_p^* \dot{\boldsymbol{\gamma}}^* \qquad (3)$$

The domain of definition of Eqs. (1)-(3) is $\{-H^*(z^*) < y^* < H^*(z^*), 0 < z^* < \ell^*\}$. The governing equations are solved with no-slip and no-penetration boundary conditions along the walls of the channel, i.e., $\mathbf{u}^*(\pm H^*(z^*), z^*) = \mathbf{0}$. Alternatively, the conditions at one of walls can be used along with symmetry conditions in the midplane, i.e. at $y^*=0$ (see below). Also, the integral constraint of mass due to fluid's incompressibility at any distance from the inlet is also utilized,

$$Q^* = \int_{-H^*(z^*)}^{H^*(z^*)} U^*(y^*, z^*) dy^* = \text{constant}$$

, and a datum pressure, $P^*_{ref}$, at the upper wall of the outlet cross-section $P^*_{ref} = P^*(H^*(\ell^*), \ell^*)$ is used. All velocity boundary conditions are provided in dimensionless scalar form below. Also, we defer for now to specify the boundary conditions for the polymer extra-stress tensor. We only mention that along the wall of the channel, Eq. (3) reduces to an algebraic equation for $\boldsymbol{\tau}^*$ (due to the fact that $\mathbf{u}^* \cdot \nabla^* \boldsymbol{\tau}^* = \mathbf{0}$) which can be solved to determine the non-trivial components of the polymer extra-stress tensor in terms of the shear-rate at the wall only.

In order to be able to apply the lubrication approximation, it is important to derive the governing equations in dimensionless form. To this end, first we define the characteristic velocity $u_c^* = Q^*/h_0^*$, and we introduce dimensionless variables by scaling $z^*$ by $\ell^*$, $y^*$ and $H^*$ by $h_0^*$, $U^*$ by $u_c^*$, and $V^*$ by $h_0^* u_c^*/\ell^*$ [37,21,45,46,51]. The pressure difference $P^* - P^*_{ref}$ is scaled by $(\eta_s^* + \eta_p^*) u_c^* \ell^* / h_0^{*2}$, and the viscoelastic extra-stress components, $\tau_{zz}^*, \tau_{yz}^*$, and $\tau_{yy}^*$ are scaled by $\eta_p^* u_c^* / h_0^*$, $\eta_p^* u_c^* \ell^* / h_0^{*2}$ and $\eta_p^* u_c^* / \ell^*$, respectively. Using these scales, the gradient operator becomes $\nabla \equiv h_0^* \nabla^* = \mathbf{e}_y (\partial / \partial y) + \varepsilon \mathbf{e}_z (\partial / \partial z)$ and the velocity vector $\mathbf{u} \equiv \mathbf{u}^* / u_c^* = \varepsilon V \mathbf{e}_y + U \mathbf{e}_z$. The rate of deformation tensor, $\dot{\gamma} \equiv h_0^* \dot{\gamma}^* / u_c^*$, polymer extra-stress tensor, $\boldsymbol{\tau} \equiv h_0^* \boldsymbol{\tau}^* / (\eta_p^* u_c^*)$, and total momentum tensor, $\mathbf{T} \equiv h_0^{*2} \mathbf{T}^* / ((\eta_s^* + \eta_p^*) u_c^* \ell^*)$, give, respectively, are:

$$\dot{\gamma} = \mathbf{e}_z \mathbf{e}_z \left(2\varepsilon \frac{\partial U}{\partial z}\right) + (\mathbf{e}_z \mathbf{e}_y + \mathbf{e}_y \mathbf{e}_z)\left(\frac{\partial U}{\partial y} + \varepsilon^2 \frac{\partial V}{\partial z}\right) + \mathbf{e}_y \mathbf{e}_y \left(2\varepsilon \frac{\partial V}{\partial y}\right) \tag{4a}$$

$$\boldsymbol{\tau} = \mathbf{e}_z \mathbf{e}_z \tau_{zz} / \varepsilon + (\mathbf{e}_z \mathbf{e}_y + \mathbf{e}_y \mathbf{e}_z) \tau_{yz} + \mathbf{e}_y \mathbf{e}_y \varepsilon \tau_{yy} \tag{4b}$$

$$\mathbf{T} = -\mathrm{Re}\,\mathbf{u}\mathbf{u} - P\mathbf{I} + \varepsilon(1-\eta)\dot{\gamma} + \varepsilon \eta \boldsymbol{\tau} \tag{4c}$$

Using the dimensionless variables, the complete set of governing equations ($\nabla \cdot \mathbf{u} = 0, \nabla \cdot \mathbf{T} = \mathbf{0}$), in scalar form, is:

$$\frac{\partial U}{\partial z} + \frac{\partial V}{\partial y} = 0 \tag{5}$$

$$\frac{\partial T_{yz}}{\partial y} + \varepsilon \frac{\partial T_{zz}}{\partial z} = 0 \tag{6a}$$

$$\varepsilon \frac{\partial T_{yz}}{\partial z} + \frac{\partial T_{yy}}{\partial y} = 0 \tag{6b}$$

where the components of the total stress tensor are:

$$T_{zz} = -\mathrm{Re}\, U^2 - P + 2(1-\eta)\varepsilon^2 \frac{\partial U}{\partial z} + \eta \tau_{zz} \tag{7a}$$

$$T_{yz} = -\mathrm{Re}\,\varepsilon UV + \varepsilon(1-\eta)\left(\frac{\partial U}{\partial y} + \varepsilon^2 \frac{\partial V}{\partial z}\right) + \eta \varepsilon \tau_{yz} \tag{7b}$$

$$T_{yy} = -\mathrm{Re}\,\varepsilon^2 V^2 - P + 2(1-\eta)\varepsilon^2 \frac{\partial V}{\partial y} + \eta \varepsilon^2 \tau_{yy} \tag{7c}$$

The non-trivial components of the constitutive model are:

$$\tau_{zz} + De\left(\frac{D\tau_{zz}}{Dt} - 2\tau_{zz}\frac{\partial U}{\partial z} - 2\tau_{yz}\frac{\partial U}{\partial y}\right) = 2\varepsilon^2 \frac{\partial U}{\partial z} \tag{8a}$$

$$\tau_{yz} + De\left(\frac{D\tau_{yz}}{Dt} - \tau_{zz}\frac{\partial V}{\partial z} - \tau_{yy}\frac{\partial U}{\partial y}\right) = \frac{\partial U}{\partial y} + \varepsilon^2 \frac{\partial V}{\partial z} \tag{8b}$$

$$\tau_{yy} + De\left(\frac{D\tau_{yy}}{Dt} - 2\tau_{yz}\frac{\partial V}{\partial z} - 2\tau_{yy}\frac{\partial V}{\partial y}\right) = 2\frac{\partial V}{\partial y} \tag{8c}$$

where $D/Dt \equiv U(\partial/\partial z) + V(\partial/\partial y)$ is the convective derivative at steady state. The dimensionless domain of definition of Eqs. (5)-(8a,b,c) is $\Omega = \{(y,z) \mid -H(z) < y < H(z), 0 < z < 1\}$. Four dimensionless numbers and parameters appear in these equations: The aspect ratio of the channel, $\varepsilon$, the Deborah number, $De$, the effective Reynolds number, $\mathrm{Re}$, and the polymer viscosity ratio, $\eta$, respectively:

$$\varepsilon \equiv \frac{h_0^*}{\ell^*}, \quad De \equiv \frac{\lambda^* u_c^*}{\ell^*}, \quad \mathrm{Re} \equiv \frac{\rho^* u_c^* h_0^{*2}}{(\eta_s^* + \eta_p^*)\ell^*}, \quad \eta \equiv \frac{\eta_p^*}{\eta_s^* + \eta_p^*} \tag{9}$$

where for the Oldroyd-B model $0 < \eta < 1$. The Deborah number is the ratio of the fluid's longest relaxation time to its characteristic residence time in the channel, while the effective Reynolds number is the product of the standard Reynolds number and the channel's aspect ratio. Note that for $\eta = 0$, Eqs. (5)-(7a,b,c) reduce to the Newtonian model, and for $\eta = 1$ to the Upper Convected Maxwell model (UCM).

The auxiliary dimensionless conditions (boundary conditions, total mass balance, symmetry conditions at the midplane, and the datum pressure) are:

$$V = U = 0 \quad \text{at} \quad y = H(z), 0 \leq z \leq 1 \tag{10a}$$

$$V = \frac{\partial U}{\partial y} = \tau_{yz} = 0 \quad \text{at} \quad y = 0, 0 \leq z \leq 1 \tag{10b}$$

$$2 \int_0^{H(z)} U(y,z) \, dy = 1 \quad \text{at} \quad 0 \leq z \leq 1 \tag{10c}$$

$$P(H(1),1) = 0 \tag{10d}$$

These conditions are augmented with the exact solution of the constitutive model along the wall of the channel, i.e., at $y = H(z)$:

$$\begin{aligned}
\tau_{zz} &= 2De\,\dot{\gamma}_w^2 + 2\varepsilon^2 H' \dot{\gamma}_w \left(-1 + De\,H'\dot{\gamma}_w\right) \\
\tau_{yz} &= \left(1 + 2De\,H'\dot{\gamma}_w\right)\dot{\gamma}_w + \varepsilon^2 H'^2 \dot{\gamma}_w \left(-1 + 2De\,H'\dot{\gamma}_w\right) \\
\tau_{yy} &= 2\left(1 + De\,H'\dot{\gamma}_w\right)H'\dot{\gamma}_w + 2De\,\varepsilon^2 \dot{\gamma}_w^2 H'^4
\end{aligned} \tag{11}$$

where $0 \leq z \leq 1$, and $\dot{\gamma}_w \equiv (\partial U/\partial y)_{y=H}$ has been used for brevity; hereafter a subscript $w$ denotes value at the wall, and the prime denotes differentiation with respect to the axial coordinate.

We emphasize that inlet/outlet conditions are not specified for any of the field variables. Although this issue requires special attention, we note that at the lubrication approximation for creeping flow, boundary conditions at the inlet and outlet of the channel are not needed for the velocity vector. As is well known, for a Newtonian fluid, specifying the velocity components along the walls is sufficient to determine the velocity field and pressure gradient along the channel. The situation is more complex for the Oldroyd-B model; however, the conditions for the polymer extra-stress tensor along the walls are fundamental in determining the flow field, the pressure gradient at each cross-section, and the average pressure drop required to drive the flow at a constant flow rate.

We proceed with the definition of the average pressure drop, $\Delta\Pi$:

$$\Delta\Pi := \int_{-H(0)}^{H(0)} P(y,0)dy - \int_{-H(1)}^{H(1)} P(y,1)dy = -\int_0^1 \frac{d}{dz}\left(\int_{-H(z)}^{H(z)} P(y,z)dy\right)dz \tag{12}$$

Applying Leibniz's rule for integrals in Eq. (12), performing integration by parts, and using the symmetry with respect to the midplane, gives:

$$\Delta\Pi = -2\int_0^1\int_0^H \frac{\partial P}{\partial z}dydz + 2\int_0^1 (H-1)\left(\frac{\partial P}{\partial z}\bigg|_{y=H} + H'\frac{\partial P}{\partial y}\bigg|_{y=H}\right)dz \tag{13}$$

Eq. (13) expresses ΔΠ in terms of the spatial pressure gradients, which can be determined from the components of the momentum balance. Notice that for a straight channel, only the first term on the right-hand side of Eq. (13) remains, while the second term arises due to the varying shape of the channel.

### 3. Lubrication Equations

For a long and narrow channel, the aspect ratio, $\varepsilon \ll 1$, is small which implies that from an asymptotic point of view, all terms in the dimensionless governing equations multiplied to $\varepsilon^2$ or $\varepsilon^4$ are much smaller compared to the other terms and can be ignored as a first approximation. This is merely corresponds to a regular perturbation scheme in terms of $\varepsilon^2$ according to which every dependent field variable $f = f(y,z)$ is expanded as $f = f_0(y,z) + \varepsilon^2 f_2(y,z) + O(\varepsilon^4)$ for $\varepsilon^2 \ll 1$. Substituting into the governing equations, expanding all quantities suitably and keeping the leading order terms, gives the final lubrication equations for an Oldroyd-B fluid:

$$\frac{\partial U_0}{\partial z} + \frac{\partial V_0}{\partial y} = 0 \tag{14}$$

$$\frac{\partial T_{zz0}}{\partial z} + \frac{\partial T_{yz0}}{\partial y} = 0 \tag{15a}$$

$$\frac{\partial T_{yy0}}{\partial y} = 0 \tag{15b}$$

where

$$T_{zz0} = -\text{Re}\,U_0^2 - P_0 + \eta\tau_{zz0} \tag{16a}$$

$$T_{yz0} = -\text{Re}\,U_0 V_0 + (1-\eta)\frac{\partial U_0}{\partial y} + \eta\tau_{yz0} \tag{16b}$$

$$T_{yy0} = -P_0 \tag{16c}$$

The lowest-order equations for the constitutive model are:

$$\tau_{zz0} + De\left(\frac{D\tau_{zz0}}{Dt} - 2\tau_{zz0}\frac{\partial U_0}{\partial z} - 2\tau_{yz0}\frac{\partial U_0}{\partial y}\right) = 0 \tag{17a}$$

$$\tau_{yz0} + De\left(\frac{D\tau_{yz0}}{Dt} - \tau_{zz0}\frac{\partial V_0}{\partial z} - \tau_{yy0}\frac{\partial U_0}{\partial y}\right) = \frac{\partial U_0}{\partial y} \tag{17b}$$

$$\tau_{yy0} + De\left(\frac{D\tau_{yy0}}{Dt} - 2\tau_{yz0}\frac{\partial V_0}{\partial z} - 2\tau_{yy0}\frac{\partial V_0}{\partial y}\right) = 2\frac{\partial V_0}{\partial y} \tag{17c}$$

Substituting Eq. (16a,b,c) into Eq. (15a,b) gives:

$$\text{Re}\left(\frac{\partial}{\partial y}(V_0 U_0) + \frac{\partial}{\partial z}(U_0^2)\right) = -\frac{dP_0}{dz} + (1-\eta)\frac{\partial^2 U_0}{\partial y^2} + \eta\left(\frac{\partial\tau_{zz0}}{\partial z} + \frac{\partial\tau_{yz0}}{\partial y}\right) \tag{18}$$

Therefore, the final lubrication equations at the classic lubrication limit are equations (14), (18), and (17a,b,c) with unknowns $U_0(y,z), V_0(y,z), P_0'(z), \tau_{zz0}(y,z), \tau_{yz0}(y,z)$ and $\tau_{yy0}(y,z)$.

The auxiliary dimensionless conditions (boundary conditions, total mass balance, symmetry conditions at the midplane, and the datum pressure) are:

$$V_0 = U_0 = 0 \quad \text{at} \quad y = H(z), 0 \le z \le 1 \tag{19a}$$

$$V_0 = \frac{\partial U_0}{\partial y} = \tau_{yz0} = 0 \quad \text{at} \quad y = 0, 0 \le z \le 1 \tag{19b}$$

$$2\int_0^{H(z)} U_0(y,z)\,dy = 1 \quad \text{at} \quad 0 \le z \le 1 \tag{19c}$$

$$P_0(z=1) = 0 \tag{19d}$$

Using $\dot{\gamma}_{0w} \equiv (\partial U_0/\partial y)_{y=H}$, the conditions for the polymer extra-stress components along the wall (at $y = H(z)$, $0 \le z \le 1$) are derived easily from Eq. (11) by omitting the $O(\varepsilon^2)$ terms:

$$\tau_{zz0} = 2De\,\dot{\gamma}_{0w}^2, \quad \tau_{yz0} = \left(1 + 2De\,H'\dot{\gamma}_{0w}\right)\dot{\gamma}_{0w}, \quad \tau_{yy0} = 2\left(1 + De\,H'\dot{\gamma}_{0w}\right)H'\dot{\gamma}_{0w} \tag{20}$$

## 4. Average pressure-drop and pressure gradient

At the classic lubrication limit, i.e., as $\varepsilon^2 \to 0$, Eq. (13) reduces to:

$$\Delta \Pi_0 = -2\int_0^1 P_0'(z)dz \qquad (21)$$

Eq. (21) shows that the average pressure-drop required to drive the flow in the channel with a constant flow rate is simply twice the pressure gradient integrated along the axial direction. Thus, we will focus solely on the determination of $P_0'(z)$ at any cross-section.

The tools used to determine the pressure gradient are global properties of the flow that must be maintained to derive physical, consistent, and accurate solutions. For steady, isothermal and incompressible flow, these properties arise from integrating the (i) mass balance, $\nabla \cdot \mathbf{u} = 0$, (ii) momentum balance, $\nabla \cdot \mathbf{T} = \mathbf{0}$, and (iii) mechanical energy balance of the flow, $\mathbf{u} \cdot (\nabla \cdot \mathbf{T}) = 0$, over the entire flow domain. Notably, applying (i) yields Eq. (19c). Additionally, the conservation of total momentum (or total force) and total mechanical energy ensures that no artificial (spurious) forces or energy are introduced—an issue that may arise, for example, from inaccuracies in the numerical solution of the governing equations. For simple Newtonian flows in the creeping flow limit, enforcing the total mass balance is sufficient to obtain the required solution. However, for complex fluids—such as generalized Newtonian fluids, where strong nonlinearities arise in the rheological model, or for viscoelastic fluids whose response is governed by hyperbolic-type equations—preserving these fundamental properties is essential. Based on these principles, we proceed with the derivation of the pressure gradient at each cross-section of the channel, which is presented in the following four subsections.

### 4.1. Pressure gradient from the momentum balance

Although we can work directly with Eq. (18), we prefer the equivalent equations (15a,b) and (16a,b,c). We integrate Eq. (15a) with respect to $y$, from the midplane to the upper wall, to find:

$$T_{yz0}(y=H) - T_{yz0}(y=0) + \int_0^{H(z)} \frac{\partial T_{zz0}}{\partial z} dy = 0 \qquad (22a)$$

Due to the symmetry with respect to the midplane $T_{yz0}(y=0) = 0$, and applying Leibniz's rule for integrals on the third term of the left-hand-side of Eq. (22a), gives:

$$T_{yz0,w} - H'T_{zz0,w} + \frac{d}{dz}\left(\int_0^{H(z)} T_{zz0}\,dy\right) = 0 \tag{22b}$$

Using Eq. (16a) and (16b) for $T_{zz0}$ and $T_{yz0}$, respectively, evaluated at the wall ($y = H(z)$), gives:

$$\left(T_{yz0} - H'T_{zz0}\right)_w = (1-\eta)\dot{\gamma}_{0w} + \eta(\tau_{yz0} - H'\tau_{zz0})_w + H'P_0 \tag{22c}$$

One sees that the quantity $(\tau_{yz0} - H'\tau_{zz0})$ at the wall is required; this can be found from Eq. (20). The latter yields $(\tau_{yz0} - H'\tau_{zz0})_w = \dot{\gamma}_{0w}$ and thus, Eq. (22c) is simplified as $\left(T_{yz0} - H'T_{zz0}\right)_w = \dot{\gamma}_{0w} + H'P_0$. The latter is substituted in Eq. (22b), along with $T_{zz0}$ from Eq. (16a), yielding:

$$\dot{\gamma}_{0w} + H'P_0 + \frac{d}{dz}\left(\int_0^{H(z)} (-\mathrm{Re}\,U_0^2 - P_0 + \eta\tau_{zz0})\,dy\right) = 0 \tag{22d}$$

Expanding the integral gives the final expression for $P_0'$ in terms of the pure viscous forces exerted by the fluid on the wall, an inertia contribution, and the contribution due to viscoelasticity of the polymer:

$$-P_0' = \frac{1}{H}\left(-\dot{\gamma}_{0w} + \mathrm{Re}\,\frac{d}{dz}\int_0^H U_0^2\,dy - \eta\frac{d}{dz}\int_0^H \tau_{zz0}\,dy\right) \tag{23}$$

At the creeping flow limit (Re=0) and neglecting fluid's viscoelasticity ($\tau_{zz0} = 0$), Eq. (23) yields the standard and well-known result that, at the lubrication limit, the pressure gradient at any cross section of the channel equals the corresponding shear rate at the wall (appropriately scaled by the shape function due to the wall variation). In the viscoelastic case, however, the above analysis reveals the importance of utilizing the exact analytical solution of the Oldroyd-B model along the walls of the channel.

### 4.2. *Pressure gradient from the total force balance*

The pressure gradient is derived here by considering the total force balance on the channel. For a Newtonian fluid (De=0) at the creeping flow limit (Re=0) with slip conditions along the wall, this has been described in Ref. [55] using the high-order (or extended) lubrication theory. Here, we neglect fluid slip along the walls and include in the analysis inertia and viscoelasticity. Specifically, starting from the momentum balance, $\nabla\cdot\mathbf{T} = \mathbf{0}$, and integrating over the fluid volume which is enclosed between the cross-section at $z = a$ (where $0 \leq a < 1$) and the cross-section at $z=1$, and applying the divergence theorem allows the conversion of the volume integral into a surface integral, leading to $\int_{\partial\Omega} \mathbf{T} \cdot \mathbf{n}\, dS = \mathbf{0}$ where $\mathbf{n}$ denotes the outward normal unit vector at almost each point of the enclosing boundary $\partial\Omega$. The boundary (surfaces) that encloses the total volume of the fluid between $z=a$ and $z=1$ and between $y=H(z)$ and $y=-H(z)$, is denoted by $\partial\Omega = \Omega_i \cup \Omega_o \cup \Omega_+ \cup \Omega_-$ where $\Omega_i$ is the cross section at $z=a$ with $\mathbf{n} = -\mathbf{e}_z$, $\Omega_o$ is the cross section at $z=1$ with $\mathbf{n} = \mathbf{e}_z$, $\Omega_+$ is the surface of the wall at $y=H(z)$ with $\mathbf{n} = (-\varepsilon H' \mathbf{e}_z + \mathbf{e}_y)/\Lambda$, and $\Omega_-$ is the surface of the wall at $y=-H(z)$ with $\mathbf{n} = (-\varepsilon H' \mathbf{e}_z - \mathbf{e}_y)/\Lambda$, where $\Lambda = \sqrt{1+\varepsilon^2 H'^2}$. Performing the dot products at each enclosing surface, and considering the symmetry with respect to the midplane, we obtain the two components of the total force balance. The component along the main flow direction is:

$$\varepsilon \int_0^{H(1)} T_{zz}(y,1)\,dy - \varepsilon \int_0^{H(a)} T_{zz}(y,a)\,dy + \int_a^1 \frac{1}{\Lambda}\left(T_{yz}(y,H) - \varepsilon H' T_{zz}(y,H)\right)dz = 0 \qquad (24a)$$

Following the perturbation expansion with respect to $\varepsilon^2$ and using the subscript to denote value at the wall, gives the lowest-order equation:

$$\int_0^{H(1)} T_{zz0}(y,1)\,dy - \int_0^{H(a)} T_{zz0}(y,a)\,dy + \int_a^1 \left(T_{yz0} - T_{zz0} H'\right)_w dz = 0 \qquad (24b)$$

The first two terms on the left-hand-side of the above equation are rearranged as follows:

$$\int_0^{H(1)} T_{zz0}(y,1)\,dy - \int_0^{H(a)} T_{zz0}(y,a)\,dy = \int_a^1 \frac{d}{dz}\left(\int_0^{H(z)} T_{zz0}(y,z)\,dy\right) dz \qquad (25a)$$

while the second term is simplified according to Eq. (22c) and the exact solution of the Oldroyd-B model along the wall, $(\tau_{yz0} - H'\tau_{zz0})_w = \dot{\gamma}_{0w}$, yielding:

$$\int_a^1 \left(T_{yz0} - T_{zz0} H'\right)_w dz = \int_a^1 \left(\dot{\gamma}_0 + H' P_0\right)_w dz \qquad (25b)$$

Substituting Eq. (25a,b) into Eq. (24b) and using Eq. (16a) gives:

$$\int_a^1 \frac{d}{dz}\left(\int_0^H \left(-\mathrm{Re}\, U_0^2 - P_0 + \eta\tau_{zz0}\right)dy\right) dz + \int_a^1 \left(\dot{\gamma}_0 + H' P_0\right)_w dz = 0 \qquad (26a)$$

Putting together the two integrals, and expanding all terms gives:

$$\int_\alpha^1 H\left(-P_0' + \frac{\dot{\gamma}_{0w}}{H} + \frac{\eta}{H}\frac{d}{dz}\int_0^H \tau_{zz0}\,dy - \frac{\mathrm{Re}}{H}\frac{d}{dz}\int_0^H U_0^2\,dy\right) dz = 0 \qquad (26b)$$

However, $H(z) \neq 0, \forall z \in [0,1]$ and since the integral holds for any $\alpha \in [0,1]$, we conclude that:

$$-P_0' + \frac{1}{H}\left(\dot{\gamma}_{0w} - \text{Re}\frac{d}{dz}\int_0^H U_0^2 dy + \eta\frac{d}{dz}\int_0^H \tau_{zz0} dy\right) = 0 \qquad (27)$$

Eq. (27) is identical to Eq. (23), formally proving that integrating the momentum balance along the main direction over a cross-section of the channel yields the same result as the total force balance on the channel. Again, the significance of the exact analytical solution of the Oldroyd-B model is evident.

### 4.3. Pressure gradient from a higher-order moment of the momentum balance

Here, we start again from the momentum balance in the z-direction (Eq. (15a)), multiply with $y^2 - H^2$, and integrate with respect to *y* from 0 to $H$:

$$\int_0^H \frac{\partial T_{zz0}}{\partial z}(y^2 - H^2) dy + \int_0^H \frac{\partial T_{yz0}}{\partial y}(y^2 - H^2) dy = 0 \qquad (28)$$

From Eq. (28), we get:

$$\int_0^H \left[\frac{\partial}{\partial z}\left((y^2 - H^2)T_{zz0}\right) + 2HH'T_{zz0}\right] dy + \int_0^H \left[\frac{\partial}{\partial y}\left((y^2 - H^2)T_{yz0}\right) - 2yT_{yz0}\right] dy = 0 \qquad (29a)$$

Expanding the integrals and applying Leibniz' rule, we find:

$$\frac{d}{dz}\int_0^H (y^2 - H^2)T_{zz0} dy + 2\int_0^H \left(HH'T_{zz0} - yT_{yz0}\right) dy = 0 \qquad (29b)$$

Substituting $T_{zz0}$ and $T_{yz0}$, as given by Eq. (16a) and (16b), respectively, into Eq. (29b), and expanding the integrals, we find:

$$\frac{2}{3}H^3 P_0' + (1-\eta)\int_0^H (y^2 - H^2)\frac{\partial^2 U_0}{\partial y^2} dy + \frac{d}{dz}\int_0^H (y^2 - H^2)(-\text{Re}U_0^2 + \eta\tau_{zz0}) dy$$
$$+ 2\int_0^H \left(\eta(HH'\tau_{zz0} - y\tau_{yz0}) - \text{Re}(HH'U_0^2 - yU_0V_0)\right) dy = 0 \qquad (29c)$$

The second term on the left-hand side can be simplified with the aid of the integral constraint of total mass, Eq. (19c). Specifically, Eq. (19c) is integrated by parts twice and applying the boundary conditions on the wall (Eq. (19a)), and the symmetry conditions in the midplane (Eq. (19b)), yields:

$$\int_0^H \left(y^2 - H^2\right)\frac{\partial^2 U_0}{\partial y^2} dy = 1 \qquad (30)$$

Substituting Eq. (30) into Eq. (29c) provides the final equation for $P_0'$:

$$-P_0' = \frac{3}{2H^3}\left\{1-\eta+\mathrm{Re}\left(2\int_0^H yU_0V_0 dy - \frac{d}{dz}\int_0^H (y^2-H^2)U_0^2 dy - 2HH'\int_0^H U_0^2 dy\right)\right. \tag{31}$$
$$\left.-\eta\left(2\int_0^H y\tau_{yz0}dy - \frac{d}{dz}\int_0^H (y^2-H^2)\tau_{zz0}dy - 2HH'\int_0^H \tau_{zz0}dy\right)\right\}$$

### 4.4. Pressure drop from the total mechanical energy

Here, we start form Eq. (18) which we multiply with the main velocity component $U_0$, and integrate from $y=0$ to $y=H$, and taking into account the total mass balance and the symmetry conditions in the midplane:

$$-P_0' = \mathrm{Re}\frac{d}{dz}\left(\int_0^H U_0^3 dy\right) + 2(1-\eta)\int_0^H \left(\frac{\partial U_0}{\partial y}\right)^2 dy + 2\eta\left[\int_0^H \left(\tau_{yz0}\frac{\partial U_0}{\partial y}+\tau_{zz0}\frac{\partial U_0}{\partial z}\right)dy - \frac{d}{dz}\left(\int_0^H (\tau_{zz0}U_0)dy\right)\right] \tag{32}$$

For a Newtonian fluid, $\tau_{zz0}=0$ and $\tau_{yz0}=\partial U_0/\partial y$, and the sum of the second and third terms on the right-hand-side of Eq. (32) give the Newtonian pure viscous contribution $2\int_0^H (\partial U_0/\partial y)^2 dy$.

Eq. (32) holds for any viscoelastic constitutive model but it can be simplified further for the Oldroyd-B model by integrating Eq. (17a) from $y=0$ to $y=H$, and taking into account the symmetry conditions to find:

$$\int_0^H \left(\tau_{yz0}\frac{\partial U_0}{\partial y}+\tau_{zz0}\frac{\partial U_0}{\partial z}\right)dy = \frac{1}{2De}\int_0^H \tau_{zz0}dy + \frac{1}{2}\frac{d}{dz}\left(\int_0^H (\tau_{zz0}U_0)dy\right) \tag{33}$$

Substituting Eq. (33) into Eq. (32) gives:

$$-P_0' = 2(1-\eta)\int_0^H \left(\frac{\partial U_0}{\partial y}\right)^2 dy + \mathrm{Re}\frac{d}{dz}\left(\int_0^H U_0^3 dy\right) - \eta\frac{d}{dz}\left(\int_0^H U_0\tau_{zz0}dy\right) + \frac{\eta}{De}\int_0^H \tau_{zz0}dy \tag{34}$$

Eq. (32) (or Eq. (34)) reveals that at the lubrication limit, $P_0'$ is decomposed into fourth contributions: the first due to viscous dissipation, the second due to fluid inertia, the third due to the work done by the elastic forces, and the fourth due to the elastic bulk contribution.

### 5. Discussion

Using the classic lubrication approximation, we derived the pressure gradient at each cross-section of a planar 2D symmetric channel with varying walls using four different methods. The first and the second methods give identical theoretical expressions, but the other two methods yield different outcomes. However, when exact or high-accuracy solutions of the equations are derived, all methods should produce the same results for the pressure gradient. Here, the main features of each method and the corresponding expressions (Eqs. (23), (27), (31) and (34)) along with their qualitative differences are discussed and compared.

The first method is based on the momentum balance at the lubrication limit, integrated directly over the cross-section of the channel. This approach requires solving the Oldroyd-B model at the wall(s), which is achieved by analytically expressing the non-trivial components of the model in terms of the shear rate at the wall(s). When this solution is used, the integrated momentum balance simplifies substantially, leading to the final expression for the pressure gradient, i.e., Eq. (23). The latter reveals three contributions to the pressure gradient: a pure viscous contribution at the wall(s), an inertial bulk contribution, and a viscoelastic bulk contribution arising exclusively from the primary normal component of the polymer extra-stress tensor. Note that this method does not impose the integral constraint due to the fluid's incompressibility explicitly. Finally, if the inertial and viscoelastic contributions are neglected—i.e., for a simple Newtonian fluid in the creeping flow limit—the pressure gradient at each cross-section equals the shear rate at the wall, $P_0' = \dot{\gamma}_{0w} / H$. This ensures that no artificial or spurious forces are introduced, as well as that the velocity field is fully consistent with the total force balance on the system. However, in the viscoelastic case, $P_0'$ must be determined through Eq. (23) in order to ensure that the conservation of the total momentum (or total force) is not violated.

The second method is based on the total force balance over the entire flow domain. Starting from the momentum balance in its divergence form, integrating over the flow domain, and applying Gauss's theorem leads to a surface integral over the boundary enclosing the flow domain. This integral requires the solution for the polymer extra-stress tensor along the walls. Taking the lubrication limit, using the exact solution for the Oldroyd-B model at the walls, and simplifying the surface integral, we obtain the general expression for the pressure gradient, Eq. (27), which fully matches the result derived from the first method. As in the first method, the constraint due to the fluid's incompressibility is not directly imposed. The previous comments remain applicable here too. Finally, note that at the creeping flow limit, the first and second methods for evaluating the pressure gradient have been used in the literature of viscoelastic fluids (see, for instance, in Refs [46,48-50]).

The third method is based on the momentum balance at the lubrication limit, first weighted by the function $f(y,z) = y^2 - H^2(z)$. The equation is then integrated over a cross section, from $y = 0$ to $y = H(z)$, and simplified using the total mass balance. This approach does not utilize the exact solution of the Oldroyd-B model along the wall(s), nor does it enforce the total force balance. Consequently, this method does not inherently prevent the introduction of spurious stresses or forces, something that can lead to overestimation or underestimation of the pressure gradient and to nonphysical results. Moreover, both the shear and primary normal extra-stress components, as well as both velocity components, appear in the final general expression for the pressure gradient (see Eq. (31)). On the other hand, when the inertia and viscoelastic bulk contributions are ignored, the pressure gradient at each cross-section reduces to its Newtonian counterpart, i.e., $P_0' = -3/(2H^3)$. For an Oldroyd-B fluid at the creeping flow limit, this method has been used in Refs. [51,52,56]. Finally, note that this method cannot be applied to generalized Newtonian fluids, as it can be confirmed by following the same procedure described in subsection 4.3.

The fourth method is based on the mechanical energy of the flow at the lubrication limit. The momentum equation is multiplied with the main velocity component and integrated with respect to *y* from $y = 0$ to $y = H(z)$. Using the constraint imposed by fluid's incompressibility and simplifying the terms related to viscoelasticity—by applying integration by parts and incorporating the integrated equation for the primary normal extra-stress tensor—yields the final expression for the pressure gradient (Eq. (34)). This expression is formulated in terms of pure viscous dissipation, viscoelastic dissipation, the work done by viscoelastic forces, and the energy associated with fluid inertia. Note that the method does not rely on the exact solution of the Oldroyd-B model along the walls, nor does it explicitly use the total force balance on the system. When the inertia contribution and viscoelastic dissipation are ignored, the pressure gradient at each cross-section equals the corresponding pure viscous dissipation energy of a Newtonian fluid, i.e., $P_0' = -2\int_0^H (\partial U_0/\partial y)^2 dy$. Using the velocity gradient found from the momentum balance and solving for $P_0'$ gives $P_0' = -3/(2H^3)$. This method, to the best of the authors' knowledge, has not yet been explored in the literature.

## 6. New set of lubrications equations

In Section 5, we presented four different methods for evaluating the pressure gradient at any cross-section of the channel, leading to three general expressions for the pressure gradient. Here, we derive a new set of lubrication equations based on a straightforward coordinate transformation that maps the varying boundaries of the flow domain onto fixed ones, the introduction of a streamfunction that inherently satisfies the continuity equation (Eq. (14)) and the total mass balance (Eq. (19c)), and a new set of polymer extra-stress tensor components obtained through a linear transformation of the components of the original extra-stress tensor. The coordinate mapping and the use of the streamfunction in the lubrication equations for the inertial flow of a simple Newtonian fluid have been introduced for the first time by Williams [18,19] for both 2D planar and 3D axisymmetric geometries. In the viscoelastic case, this mapping in conjunction with the streamfunction has extensively been described and applied by Housiadas & Beris [46,50]. The new polymer extra-stress tensor components are introduced here for the first time.

In particular, the streamfunction $\Psi$ is defined with the aid of $U_0$ and $V_0$:

$$U_0 = \frac{\partial \Psi}{\partial y}, \quad V_0 = -\frac{\partial \Psi}{\partial z} \tag{35}$$

Additionally, the new independent coordinates $(Y, Z)$ that map the varying boundaries of the flow domain into fixed ones are:

$$Y = \frac{y}{H(z)}, \quad Z = z \tag{36}$$

In terms of the new coordinates, the domain of definition of the lubrication equations becomes $\{(Y, Z) | -1 < Y < 1, 0 < Z \leq 1\}$. Note that Williams [18,19] found solutions to the lubrication equations for a Newtonian fluid that do not depend on the transformed axial coordinate $Z$; in this case, $Y$ is referred to as the similarity variable. In the viscoelastic case, at the creeping flow limit and for a hyperbolic 2D planar channel or a hyperbolic 3D axisymmetric pipe, similarity solutions for the Oldroyd-B model have been derived in Ref. [53].

Based on Eq. (36), the original differential operators and the material derivative are transformed as follows:

$$\frac{\partial}{\partial y} = \frac{1}{H}\frac{\partial}{\partial Y}, \quad \frac{\partial}{\partial z} = \frac{\partial}{\partial Z} - \frac{Y H'(Z)}{H(Z)}\frac{\partial}{\partial Y}, \quad \frac{D}{Dt} \equiv \frac{1}{H}\left(\frac{\partial \Psi}{\partial Y}\frac{\partial}{\partial Z} - \frac{\partial \Psi}{\partial Z}\frac{\partial}{\partial Y}\right) \tag{37}$$

while higher order derivatives can be obtained from Eq. (37). Finally, we introduce $\tau_{11}$, $\tau_{12}$, and $\tau_{22}$, according to the linear expressions:

$$\tau_{11} = \tau_{zz0}, \quad \tau_{12} = \tau_{yz0} - Y H' \tau_{zz0}, \quad \tau_{22} = \tau_{yy0} - 2Y H' \tau_{yz0} + Y^2 H'^2 \tau_{zz0} \qquad (38)$$

Eq. (38) can be inverted to uniquely determine $\tau_{zz0}$, $\tau_{yz0}$, and $\tau_{yy0}$ in terms of $\tau_{11}$, $\tau_{12}$, and $\tau_{22}$:

$$\tau_{zz0} = \tau_{11}, \quad \tau_{yz0} = \tau_{12} + Y H' \tau_{11}, \quad \tau_{yy0} = \tau_{22} + 2Y H' \tau_{12} + Y^2 H'^2 \tau_{11} \qquad (39)$$

The generalization of Eq. (38) and its inverse transformation (Eq. (39)) has been described in detail by Housiadas *et al.* 2025 [57].

Using Eqs. (35), (36) and (39) into Eqs. (15a)-(17c) and after straightforward algebraic manipulations gives the final system of partial differential equations for $\Psi, \tau_{11}, \tau_{12}$ and $\tau_{22}$:

$$\text{Re}\left(\left(\frac{\partial \Psi}{\partial Y}\frac{\partial^2 \Psi}{\partial Y \partial Z} - \frac{\partial^2 \Psi}{\partial Y^2}\frac{\partial \Psi}{\partial Z}\right)H - \left(\frac{\partial \Psi}{\partial Y}\right)^2 H'\right) = -\frac{dP_0}{dZ}H^3 + (1-\eta)\frac{\partial^3 \Psi}{\partial Y^3} + \eta H^2\left(\frac{\partial \tau_{12}}{\partial Y} + \frac{\partial (H\tau_{11})}{\partial Z}\right) \qquad (40)$$

$$\tau_{11} + De\left(\frac{D\tau_{11}}{Dt} - \frac{2\tau_{11}}{H}\left(\frac{\partial^2 \Psi}{\partial Y \partial Z} - \frac{H'}{H}\frac{\partial \Psi}{\partial Y}\right) - \frac{2\tau_{12}}{H^2}\frac{\partial^2 \Psi}{\partial Y^2}\right) = 0 \qquad (41)$$

$$\tau_{12} + De\left(\frac{D\tau_{12}}{Dt} - \tau_{11}\left(\frac{H'}{H}\frac{\partial \Psi}{\partial Z} - \frac{\partial^2 \Psi}{\partial Z^2}\right) - \frac{\tau_{22}}{H^2}\frac{\partial^2 \Psi}{\partial Y^2}\right) = \frac{1}{H^2}\frac{\partial^2 \Psi}{\partial Y^2} \qquad (42)$$

$$\tau_{22} + De\left(\frac{D\tau_{22}}{Dt} - 2\tau_{12}\left(\frac{H'}{H}\frac{\partial \Psi}{\partial Z} - \frac{\partial^2 \Psi}{\partial Z^2}\right) + \frac{2\tau_{22}}{H}\left(\frac{\partial^2 \Psi}{\partial Y \partial Z} - \frac{H'}{H}\frac{\partial \Psi}{\partial Y}\right)\right) = -\frac{2}{H}\left(\frac{\partial^2 \Psi}{\partial Y \partial Z} - \frac{H'}{H}\frac{\partial \Psi}{\partial Y}\right) \qquad (43)$$

Differentiating Eq. (40) with respect to *Y* eliminates the modified pressure gradient term (the first term on the right-hand-side of Eq. (40)) and gives a fourth order differential partial differential equation for the streamfunction:

$$\text{Re}\left(\left(\frac{\partial \Psi}{\partial Y}\frac{\partial^3 \Psi}{\partial Z \partial Y^2} - \frac{\partial^3 \Psi}{\partial Y^3}\frac{\partial \Psi}{\partial Z}\right)H - 2\frac{\partial \Psi}{\partial Y}\frac{\partial^2 \Psi}{\partial Y^2}H'\right) = (1-\eta)\frac{\partial^4 \Psi}{\partial Y^4} + \eta H^2\left(\frac{\partial^2 \tau_{12}}{\partial Y^2} + \frac{\partial^2 (H\tau_{11})}{\partial Y \partial Z}\right) \qquad (44)$$

Due to the appearance of $\partial^4 \Psi / \partial Y^4$, the equations must be solved with the accompanied boundary conditions for $\Psi$ at the wall ($Y=1$) and the midplane ($Y=0$), valid for $0 \leq Z \leq 1$:

$$\Psi(1,Z) = \frac{1}{2}, \quad \Psi(0,Z) = \frac{\partial^2 \Psi}{\partial Y^2}(0,Z) = \frac{\partial \Psi}{\partial Y}(1,Z) = 0 \qquad (45)$$

The exact solution of the extra-stresses at the wall ($Y=1$) is:

$$\tau_{11}(1,Z) = 2De\dot{\gamma}_{0w}^2, \quad \tau_{12}(1,Z) = \dot{\gamma}_{0w}, \quad \tau_{22}(1,Z) = 0 \qquad (46)$$

where $\dot{\gamma}_{0w} = \partial U_0 / \partial y\big|_{y=H} = H^{-2}(\partial^2 \Psi / \partial Y^2)\big|_{Y=1}$. It is also worth mentioning that the exact solution of the Oldroyd-B model along the wall (Eq. (46)), expressed in terms of the mapped

coordinates and the new components of the extra-stress tensor, is simpler than the solution given by Eq. (20) in terms of the original coordinates and the components of the original extra-stress tensor.

The new form of Eqs. (23), (31) and (34) are, respectively:

$$-P_0' = -\frac{1}{H^3}\left(\frac{\partial^2 \Psi}{\partial Y^2}\right)_w + \frac{1}{H}\frac{d}{dZ}\left(\text{Re}\frac{I}{H} - \eta H E\right)$$

$$I = \int_0^1 \left(\frac{\partial \Psi}{\partial Y}\right)^2 dY, \quad E = \int_0^1 \tau_{11} dY$$

(47)

$$-P_0' = \frac{3(1-\eta)}{2H^3} + \frac{3\text{Re}}{2H}\hat{I} - \frac{\eta}{H}\hat{E}$$

$$\hat{I} = \frac{d}{dZ}\left(\frac{1}{H}\int_0^1 (1-Y^2)\left(\frac{\partial \Psi}{\partial Y}\right)^2 dY\right) - \frac{2}{H}\int_0^1 Y \frac{\partial \Psi}{\partial Y}\frac{\partial \Psi}{\partial Z} dY$$

$$\hat{E} = \int_0^1 \left(\frac{3}{2}(1-Y^2)\frac{d}{dZ}(H\tau_{11}) + 3Y\tau_{12}\right) dY$$

(48)

$$-P_0' = \frac{2(1-\eta)}{H^3}\int_0^1 \left(\frac{\partial^2 \Psi}{\partial Y^2}\right)^2 dY + \text{Re}\frac{d\tilde{I}}{dZ} + \eta\left(-\frac{d\tilde{E}}{dZ} + \frac{H}{De}\int_0^1 \tau_{11} dY\right)$$

$$\tilde{I} = \frac{1}{H^2}\int_0^1 \left(\frac{\partial \Psi}{\partial Y}\right)^3 dY, \quad \tilde{E} = \int_0^1 \frac{\partial \Psi}{\partial Y}\tau_{11} dY$$

(49)

Eq. (47) represents the total force balance on the system, Eq. (48) is the higher-order moment of the momentum balance, and Eq. (49) describes the total mechanical energy of the flow. However, within the streamfunction formulation, Eqs. (48) and (49) do not explicitly satisfy the total force balance and therefore do not offer any advantages over Eq. (47).

The new set of lubrication equations is well-suited for both numerical and analytical purposes. A key advantage of this formulation is that mass fluxes, namely the continuity equation, and the total mass balance are satisfied exactly due to the use of the streamfunction. Furthermore, once the streamfunction is determined, the pressure gradient follows directly from Eq. (47), since among Eqs. (47)-(49), is the simplest and easiest to implement. Consequently, any method—numerical, asymptotic, or analytic—that solves Eqs. (41)–(44), along with the appropriate boundary conditions (such as those in Eqs. (45)–(46)) and initial conditions and determines the pressure-gradient from Eq. (47) avoids errors associated with the enforcement of the continuity equation, the total mass balance, and the total force balance.

### 7. A nonlinear case: Inertial Newtonian flow in a linear channel

As a demonstration of the equivalence of the theoretical formulas for the pressure gradient at any cross-section of the channel and the average pressure drop required to maintain a constant flow rate through the channel, we present the case of Newtonian inertia flow in a linearly varying channel. This flow was first investigated by Williams [18,19], although he did not provide an evaluation of the pressure gradient, or of the average pressure drop.

For the linear channel the shape function is:

$$H(Z) = 1 + \left(\frac{1}{\Lambda} - 1\right)Z = 1 + \beta Z \qquad (50)$$

where $\Lambda = h_0^* / h_f^*$ is the contraction ratio, and $\beta \equiv H'(Z) = 1/\Lambda - 1$ is used for brevity. Ignoring the viscoelastic contribution ($\eta=0$), Eq. (40) reduces to:

$$\mathrm{Re}\left(\left(\frac{\partial \Psi}{\partial Y}\frac{\partial^2 \Psi}{\partial Y \partial Z} - \frac{\partial^2 \Psi}{\partial Y^2}\frac{\partial \Psi}{\partial Z}\right)H - \left(\frac{\partial \Psi}{\partial Y}\right)^2 H'\right) = G + \frac{\partial^3 \Psi}{\partial Y^3} \qquad (51)$$

where $G = G(Z) \equiv -P_0'(Z)H^3(Z)$ is found by multiplying Eqs. (47)-(49) with $H^3$ and setting $\eta=0$:

$$G = -\left.\frac{\partial^2 \Psi}{\partial Y^2}\right|_{Y=1} + \mathrm{Re}\, H^2 \frac{d}{dZ}\left(\frac{1}{H}\int_0^1 \left(\frac{\partial \Psi}{\partial Y}\right)^2 dY\right) \qquad (52a)$$

$$G = \frac{3}{2} + \frac{3\mathrm{Re}}{2}H^2\left\{\frac{d}{dZ}\left(\frac{1}{H}\int_0^1 (1-Y^2)\left(\frac{\partial \Psi}{\partial Y}\right)^2 dY\right) - \frac{2}{H}\int_0^1 Y \frac{\partial \Psi}{\partial Y}\frac{\partial \Psi}{\partial Z}dY\right\} \qquad (52b)$$

$$G = 2\int_0^1 \left(\frac{\partial^2 \Psi}{\partial Y^2}\right)^2 dY + \mathrm{Re}\, H^3 \frac{d}{dZ}\left(\frac{1}{H^2}\int_0^1 \left(\frac{\partial \Psi}{\partial Y}\right)^3 dY\right) \qquad (52c)$$

We will be referring to $G$ as the *modified pressure gradient* at any cross-section of the channel. Similarly, Eq. (44) gives:

$$\mathrm{Re}\left(\left(\frac{\partial \Psi}{\partial Y}\frac{\partial^3 \Psi}{\partial Z \partial Y^2} - \frac{\partial^3 \Psi}{\partial Y^3}\frac{\partial \Psi}{\partial Z}\right)H - 2\frac{\partial \Psi}{\partial Y}\frac{\partial^2 \Psi}{\partial Y^2}H'\right) = \frac{\partial^4 \Psi}{\partial Y^4} \qquad (53)$$

For creeping flow (Re=0), and using the subscript "*cr*", the solutions for the streamfunction and the modified pressure gradient can be obtained very easily:

$$\Psi_{cr}(Y) = \frac{3Y}{4} - \frac{Y^3}{4}, \quad G_{cr} = \frac{3}{2} \qquad (54)$$

This is a Poiseuille-type solution for the flow variables in the channel. Notice that both the streamfunction and the pressure gradient do not depend on the transformed axial coordinate (recall that $Y \equiv y/H(z)$ and $Z \equiv z$).

For Re > 0, the partial differential equation (PDE) governing the spatial evolution of the streamfunction (i.e., Eq. (51)), along with the accompanying boundary conditions (i.e., Eq. (45)) and a fully developed Poiseuille velocity profile at the inlet of the varying section of the channel, $\Psi(Y, Z=0) = \Psi_{cr}(Y)$, has been solved by Sialmas & Housiadas [58] using high-accuracy methods. In particular, these authors developed numerical methods of spectral accuracy, as well as asymptotic analytical techniques, building on the successful approaches previously introduced by Housiadas & Beris [46,50] and Sialmas & Housiadas [53], to improve the accuracy, stability, and efficiency of the computations. The simulations were performed over a wide range of contraction ratios and Reynolds numbers. Notably, all computations were carried out in a fully implicit manner using a Newton iterative scheme with an absolute convergence criterion of 10$^{-14}$. In their earlier work [53], Sialmas & Housiadas employed the first theoretical expression to calculate the pressure gradient, namely Eq. (47). In the present study, we repeat the calculations using all theoretical expressions, Eqs. (47)–(49), to demonstrate their theoretical and numerical equivalence.

In the following, we refer to the spectrally accurate version of our code that solves Eq. (51) as Version A, and to the version that solves Eq. (53) as Version B. Specifically, Version A-1 solves the third-order PDE (Eq. (51)) along with the boundary conditions given in Eq. (45) and the initial condition $\Psi(Y, Z=0) = \Psi_{cr}(Y)$, to calculate $\Psi = \Psi(Y, Z)$. At each cross-section of the channel, $G$ is determined from Eq. (52a) simultaneously with the streamfunction, in a fully coupled manner. Version A-2 performs the same calculations, but $G$ is instead determined from Eq. (52b). Finally, Version A-3 follows the same procedure as Versions A-1 and A-2, but determines $G$ using Eq. (52c).

Version B of the spectral code solves the fourth-order PDE (Eq. (53)) along with the boundary conditions given in Eq. (45) and the initial condition $\Psi(Y, Z=0) = \Psi_{cr}(Y)$ to calculate $\Psi = \Psi(Y, Z)$ numerically. However, at each cross-section of the channel, $G$ is determined *a posteriori*—after obtaining the streamfunction—using all available theoretical expressions, i.e., Eq. (52a), (52b), and (52c).

The agreement in $G$ obtained from Versions A-1, A-2, A-3, and B is accurate to seven to ten significant digits. This clearly demonstrates that, when the velocity field is resolved with high accuracy using fully implicit schemes, Eqs. (52a–c) are both theoretically and numerically equivalent expressions for the modified pressure gradient.

Although this nonlinear case and the corresponding numerical results provide strong evidence for the validity and accuracy of the theoretical methods presented here, we take an additional step by considering a similarity solution of Eq. (51) (or Eq. (53)). A similarity solution is a powerful tool for solving differential equations, particularly in the context of nonlinear PDEs [59]. It typically reduces the number of independent variables (i.e., spatial and/or temporal coordinates) by combining them into a single similarity variable, thereby simplifying the governing equation(s) significantly. Such solutions often capture the fundamental behavior of the system and may represent asymptotic states that more general solutions tend toward. Despite their many advantages, similarity solutions do not satisfy all possible initial and boundary conditions—a key limitation. In general, they satisfy only a subset of conditions that are compatible with the similarity form, specifically those exhibiting self-similar structure.

### 7.1. Similarity solution

For the linearly varying channel, Eq. (51), or Eq. (53), admits a similarity solution, as first identified and derived by Williams [18,19]. Particularly, assuming that $\Psi(Y,Z) \approx \hat{\Psi}(Y)$ and $G(Z) \approx \hat{G} = \text{constant}$, Eqs. (51) and (53) reduce to:

$$\text{Re}\,\beta\left(\hat{\Psi}'(Y)\right)^2 + \hat{G} + \hat{\Psi}'''(Y) = 0 \tag{55a}$$

and

$$2\,\text{Re}\,\beta\,\hat{\Psi}'(Y)\hat{\Psi}''(Y) + \hat{\Psi}^{(4)}(Y) = 0 \tag{55b}$$

Eq. (55a) is a third-order ordinary differential equation (ODE) for $\hat{\Psi}$ which contains the unknown modified pressure gradient $\hat{G}$, whereas Eq. (55b) is a fourth-order ODE for $\hat{\Psi}$ which does contain $\hat{G}$. Since Eqs. (55a) and (55b) are ODEs, an initial condition at Z=0 cannot be applied. Therefore, the similarity solution for the streamfunction is a special solution of the original lubrication equation (Eqs. (51) or (53)) which satisfies all boundary conditions at the walls of the channel and the symmetry conditions on the midplane, but not the initial condition at the inlet.

Eq. (55a), or Eq. (55b), is solved with the fully spectral method developed by Sialmas & Housiadas [53,58] along with the boundary conditions given in Eq. (45). These boundary conditions in terms of $\hat{\Psi}$ are:

$$\hat{\Psi}(1) = 1/2, \quad \hat{\Psi}(0) = \hat{\Psi}''(0) = \hat{\Psi}'(1) = 0 \tag{56}$$

It is worth mentioning that the similarity solution at the creeping flow limit is:

$$\hat{\Psi}_{cr}(Y) = \frac{3Y}{4} - \frac{Y^3}{4}, \quad \hat{G}_{cr} = \frac{3}{2} \tag{57}$$

Comparing Eq. (57) with Eq. (54) shows that for creeping flow, the full and similarity solutions are identical.

For the modified pressure gradient $\hat{G}$, Eqs. (52a,b,c) simplify to:

$$\hat{G} = -\hat{\Psi}''(1) - \operatorname{Re}\beta \int_0^1 \left(\hat{\Psi}'(Y)\right)^2 dY \tag{58a}$$

$$\hat{G} = \frac{3}{2}\left(1 - \operatorname{Re}\beta \int_0^1 (1-Y^2)\left(\hat{\Psi}'(Y)\right)^2 dY\right) \tag{58b}$$

$$\hat{G} = 2\int_0^1 \left(\hat{\Psi}''(Y)\right)^2 dY - 2\operatorname{Re}\beta \int_0^1 \left(\hat{\Psi}'(Y)\right)^3 dY \tag{58c}$$

According to the fully spectral method, $\hat{\Psi}$ is written as a truncated series of $N$ Legendre odd orthogonal polynomials $\hat{\Psi}(Y) = \sum_{k=0}^{N} a_{2k+1} L_{2k+1}(Y)$, where $L_{2k+1} = L_{2k+1}(Y)$ is the Legendre polynomial of degree 2$k$+1, and $a_{2k+1}$ is the corresponding spectral coefficient. The spectral coefficients are computed using a Galerkin-type approach: Eq. (55a) or Eq. (55b) is first weighted by the appropriate Legendre polynomials, and the resulting equations are integrated from $Y$=−1 to $Y$=1. The integration is performed symbolically using the *MATHEMATICA* software [60], thereby avoiding truncation and round-off errors, and yielding a nonlinear algebraic system of equations with the spectral coefficients of the streamfunction as unknowns. Solving this algebraic system completes the solution.

For Λ = 5 and Re = 1, 10, and 20, the magnitude of the spectral coefficients is shown in Figure 2. The results clearly demonstrate that the streamfunction is resolved down to machine accuracy. The linear decrease of the coefficients in the log-linear scale is evident, indicating an exponential drop and confirming the spectral accuracy of the method.

Additionally, we confirm that the differences between the spectral coefficients computed by solving Eq. (55a) or (55b) are less than 10⁻¹⁴ (i.e., close to close to machine accuracy).

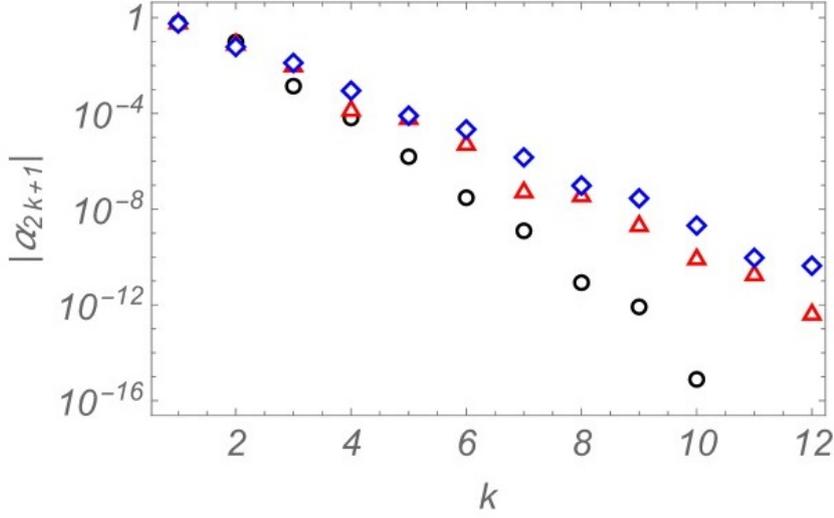

**Figure 2:** The magnitude of the spectral coefficients for the similarity solution for the streamfunction at Re = 1 (circles), Re = 10 (triangles), and Re = 20 (squares).

### 7.2. Comparison of the full and similarity solutions

Here we compare the reduced average pressure-drop $\Delta\Pi/\Delta\Pi_{cr}$, where $\Delta\Pi_{cr} \equiv \Delta\Pi(\mathrm{Re}=0)$, as $\Delta\Pi$ is computed using the numerical solution of Eq. (51) (or Eq. (53)) and the corresponding similarity solution. It is important to reiterate that the full equation is solved using the initial condition of a fully developed Poiseuille flow at the inlet, whereas the similarity solution cannot satisfy any initial condition.

Substituting Eqs. (47)–(49) into Eq. (21) and integrating with respect to Z, we derive the following expressions for $\Delta\Pi$:

$$\Delta\Pi = -2\int_0^1 \frac{1}{H^3}\frac{\partial^2\Psi}{\partial Y^2}\bigg|_{Y=1} dZ + 2\mathrm{Re}\int_0^1\left(\frac{1}{H}\frac{d}{dZ}\left(\frac{1}{H}\int_0^1\left(\frac{\partial\Psi}{\partial Y}\right)^2 dY\right)\right)dZ \quad (59a)$$

$$\Delta\Pi = 3\int_0^1 \frac{dZ}{H^3} + 3\mathrm{Re}\int_0^1 \frac{1}{H}\left(\frac{d}{dZ}\left(\frac{1}{H}\int_0^1(1-Y^2)\left(\frac{\partial\Psi}{\partial Y}\right)^2 dY\right) - \frac{2}{H}\int_0^1 Y\frac{\partial\Psi}{\partial Y}\frac{\partial\Psi}{\partial Z}dY\right)dZ \quad (59b)$$

$$\Delta\Pi = 4\int_0^1 \frac{1}{H^3}\left(\int_0^1\left(\frac{\partial^2\Psi}{\partial Y^2}\right)^2 dY\right)dZ - 2\mathrm{Re}\,\Delta\left\{\frac{1}{H^2}\int_0^1\left(\frac{\partial\Psi}{\partial Y}\right)^3 dY\right\} \quad (59c)$$

where $\Delta f := f(Z=0) - f(Z=1)$ for any dependent field variable. For Re=0, Eqs. (59a-c) give $\Delta \Pi_{cr} = 3\Lambda(\Lambda+1)/2$.

The expression for $\Delta \Pi_{cr}$ holds for the similarity solution too. Additionally, since for creeping flow $\hat{G}_{cr} = 3/2$ (see Eq. (57b)), the reduced average pressure-drop which results from the similarity solution is:

$$\frac{\Delta \Pi}{\Delta \Pi_{cr}} = \frac{2}{3}\hat{G} \qquad (60)$$

In Table I, we present $\Delta \Pi / \Delta \Pi_{cr}$ for various values of the Reynolds numbers for Λ=5, as obtained by solving Eq. (51) or Eq. (53)—indicated as the "full solution" as well as by solving Eq. (55a) or (55b)—indicated as the "similarity solution". The agreement of the results demonstrates the excellent consistency between the full equation and the similarity equation. Notice that the agreement diminishes at a very slow rate as the Reynolds number increases. Similar results are also shown in Table 2 for the same values of the Reynolds number and Λ=10.

| Re | 0.1 | 1 | 5 | 10 | 20 |
|---|---|---|---|---|---|
| Full solution | 1.020 | 1.203 | 1.976 | 2.882 | 4.583 |
| Similarity solution | 1.020 | 1.203 | 1.979 | 2.893 | 4.617 |

**Table 1**: The results for $\Delta \Pi / \Delta \Pi_{cr}$, for the linear channel with Λ=5.

| Re | 0.1 | 1 | 5 | 10 | 20 |
|---|---|---|---|---|---|
| Full solution | 1.023 | 1.228 | 2.095 | 3.111 | 5.024 |
| Similarity solution | 1.023 | 1.228 | 2.096 | 3.114 | 5.034 |

**Table 2**: The results for $\Delta \Pi / \Delta \Pi_{cr}$, for the linear channel with Λ=10.

## 8. Conclusions

In the classic lubrication approximation for confined flows, the momentum balance equations are significantly simplified, as many terms—such as second derivatives of the velocity components with respect to the axial coordinate—are neglected. Consequently, for inertialess flow, entry and exit effects (i.e., the influence of velocity boundary conditions) cannot be accounted for and are completely ignored, while for inertial flow, only an initial condition for the main velocity component can be imposed. Most notably, the pressure gradient depends solely on the axial coordinate, requiring its evaluation at each cross-section of the channel.

In this work, we derived three fundamental theoretical expressions for evaluating the pressure gradient in viscoelastic inertial flow. Additionally, we developed a new set of lubrication equations that, by construction exactly satisfy the continuity equation and the total mass balance and determines the pressure gradient only after solving for the streamfunction, ensuring consistency with (i) the exact analytical solution of the polymer extra-stress tensor along the channel walls, and (ii) the total momentum balance of the flow. These features are essential for obtaining physically consistent solutions that uphold the fundamental conservation principles of mass, momentum, and energy in the flow system.

The theoretical and numerical equivalence of the general expressions derived for the pressure gradient and the average pressure drop required to maintain a constant volumetric flow rate in the channel was demonstrated for a limiting nonlinear case. Specifically, the Newtonian inertia flow in a linearly varying channel was considered. The final equation for the streamfunction, along with the appropriate boundary and initial conditions, was solved using high-accuracy spectral numerical methods, which resolved the streamfunction down to machine accuracy. All theoretical expressions for the pressure gradient were utilized, yielding numerical results that agreed within seven to ten significant digits, thus confirming the robustness and correctness of our theoretical developments and numerical algorithms.

Finally, for the limiting nonlinear case, a similarity solution for the streamfunction was also presented. This special solution satisfies all boundary and symmetry conditions, but not the initial condition used to solve the full equation. Despite this limitation, the agreement between the full and similarity solutions was excellent, highlighting the invaluable role of similarity solutions in PDEs that model transport phenomena, such as flows in confined and slender geometries.

**APPENDIX**

For creeping flow, the new set of lubrication equations derived in Section 6 can be shown to be identical to those derived by Hinch *et al.* [51] using orthogonal curvilinear coordinates and different methods and techniques than those employed in the present work. First, we introduce the viscosity ratio $c = \eta/(1-\eta) = \eta_p^*/\eta_s^*$ and redefine the pressure variable as $p = P_0/(1-\eta)$. Additionally, we define the new velocity components:

$$u = \frac{1}{H}\frac{\partial \Psi}{\partial Y}, \quad v = -\frac{\partial \Psi}{\partial Z} \tag{A1}$$

which satisfy the equation:

$$\frac{\partial(uH)}{\partial Z} + \frac{\partial v}{\partial Y} = 0 \tag{A2}$$

Using $u$ and $v$, the material derivative becomes:

$$\frac{D}{Dt} = \frac{1}{H}\frac{\partial \Psi}{\partial Y}\frac{\partial}{\partial Z} - \frac{1}{H}\frac{\partial \Psi}{\partial Z}\frac{\partial}{\partial Y} = u\frac{\partial}{\partial Z} + \frac{v}{H}\frac{\partial}{\partial Y} \tag{A3}$$

Also, we define the velocity gradients:

$$\gamma_1 = \frac{1}{H^2}\frac{\partial^2 \Psi}{\partial Y^2} = \frac{1}{H}\frac{\partial u}{\partial Y}, \quad \gamma_2 = H\frac{\partial}{\partial Z}\left(-\frac{1}{H}\frac{\partial \Psi}{\partial Z}\right) = H\frac{\partial}{\partial Z}\left(\frac{v}{H}\right), \quad e = \frac{1}{H}\frac{\partial^2 \Psi}{\partial Y \partial Z} - \frac{H'}{H^2}\frac{\partial \Psi}{\partial Y} = \frac{\partial u}{\partial Z} \tag{A4}$$

Finally, we consider the components of the conformation tensor at the lubrication limit:

$$\tau_{11} = \frac{A_{11}}{De}, \quad \tau_{12} = \frac{A_{12}}{De}, \quad \tau_{22} = \frac{A_{22}-1}{De} \tag{A5}$$

Using (A1)-(A5) in Eq. (40), we find:

$$\text{Re}\frac{Du}{Dt} = -\frac{dp}{dZ} + \frac{1}{H^2}\frac{\partial^2 u}{\partial Y^2} + \frac{c}{De}\left(\frac{1}{H}\frac{\partial(HA_{11})}{\partial Z} + \frac{1}{H}\frac{\partial A_{12}}{\partial Y}\right) \tag{A6}$$

Finally, from Eqs (41)-(43), we find:

$$\frac{DA_{11}}{Dt} - 2eA_{11} - 2\gamma_1 A_{12} = -\frac{A_{11}}{De} \tag{A7a}$$

$$\frac{DA_{12}}{Dt} - \gamma_2 A_{11} - \gamma_1 A_{22} = -\frac{A_{12}}{De} \tag{A7b}$$

$$\frac{DA_{22}}{Dt} - 2\gamma_2 A_{12} + 2eA_{22} = -\frac{A_{22}-1}{De} \tag{A7c}$$

For Re=0, Eqs. (A2), (A6), (A7a), (A7b), and (A7c) are identical to (2.1) and (2.2), (2.3a), (2.3b), and (2.3c), respectively, derived by Hinch *et al.* [51]. Finally, substituting Eq. (A5) into Eq. (48) reduces to:

$$\frac{dp}{dZ} = -\frac{3}{2H^3} + \frac{c}{H\,De}\int_0^1 \left(\frac{3}{2}(1-Y^2)\frac{d}{dZ}(HA_{11}) + 3Y A_{12}\right)dY \qquad (A8)$$

Eq. (A8) is identical to Eq. (2.8) derived and used in Ref. [51]; the additional coefficient ½ that appears on the first term of the right-hand-side of Eq. (A8) is because the dimensionless total flux used in this work is unity, while in Ref. [51] the total flux is unity in the half of the cross-section of the channel. Thus, these authors chose to compute the pressure gradient using the third method presented here—i.e., based on the higher-order moment of the momentum balance—while ignoring the total force balance and without mentioning whether the exact solution of the Oldroyd-B model along the walls is satisfied precisely. As explained in the main text—and clearly demonstrated for the nonlinear flow problem in Section 7—when the field variables are resolved down to machine accuracy, all approaches for computing the pressure gradient yield almost identical numerical results. However, low accuracy numerical methods or low accuracy simulations, do not guarantee the avoidance of spurious stresses (or forces) and could potentially lead to a significant underestimation or overestimation of the pressure gradient, particularly at high Deborah numbers.